\documentclass{article}

\usepackage{PRIMEarxiv}

\usepackage[utf8]{inputenc} % allow utf-8 input
\usepackage[T1]{fontenc}    % use 8-bit T1 fonts
\usepackage{hyperref}       % hyperlinks
\usepackage{url}            % simple URL typesetting
\usepackage{booktabs}       % professional-quality tables
\usepackage{amsfonts}       % blackboard math symbols
\usepackage{nicefrac}       % compact symbols for 1/2, etc.
\usepackage{microtype}      % microtypography
\usepackage{lipsum}
\usepackage{fancyhdr}       % header
\usepackage{graphicx}       % graphics
\graphicspath{{media/}}     % organize your images and other figures under media/ folder
\usepackage[version=4]{mhchem}
\usepackage{xcolor}
\title{Hidden Defect Chemistry in Ion-Irradiated MoS$_2$ Field-Effect Transistors Revealed by Photocurrent Loss}
\author{Leon Daniel \\
  Faculty of Physics and CENIDE \\
  University of Duisburg-Essen \\
  Duisburg\\
   \\
   \And
   Dedi Sutarma\\ 
  Faculty of Physics and CENIDE \\
  University of Duisburg-Essen \\
  Duisburg\\
     \And
   Leon Klieve\\ 
  Faculty of Physics and CENIDE \\
  University of Duisburg-Essen \\
  Duisburg\\
     \And
  Osamah Kharsah\\ 
  Faculty of Physics and CENIDE \\
  University of Duisburg-Essen \\
  Duisburg\\
     \And
   Ulrich Hagemann\\ 
  ICAN and CENIDE \\
  University of Duisburg-Essen \\
  Duisburg\\
     \And
  Oliver Altenhoff\\ 
  Faculty of Physics and CENIDE \\
  University of Duisburg-Essen \\
  Duisburg\\
       \And
  Stephan Sleziona\\ 
  Faculty of Physics and CENIDE \\
  University of Duisburg-Essen \\
  Duisburg\\
  \And
   Lars Breuer\\ 
  Faculty of Physics and CENIDE \\
  University of Duisburg-Essen \\
  Duisburg\\
       \And
        Peter Kratzer\\ 
  Faculty of Physics and CENIDE \\
  University of Duisburg-Essen \\
  Duisburg\\
       \And
  Marika Schleberger*\\ 
  Faculty of Physics and CENIDE \\
  University of Duisburg-Essen \\
  Duisburg\\
  \texttt{marika.schleberger@uni-due.de} \\
}

\begin{document}
\maketitle
\begin{abstract}
Defect engineering in monolayer \ce{MoS2} is a promising route to tune field-effect transistors (FETs), but the electronic response of defects in processed devices can be masked by contacts, substrate effects, adsorbates, and chemical passivation. Here, we irradiate MoS$_2$ FETs with low-energy 40~eV Ar$^+$ ions to preferentially create sulfur vacancies (V\textsubscript{S}) in the channel while minimizing substrate damage. We compare dark and illuminated electrical characterization with surface analysis and first-principles calculations. Dark transfer characteristics show an apparent robustness against irradiation up to moderate fluences, with pronounced degradation only at the highest fluence. Under 532~nm illumination, however, the photocurrent and light-induced photodoping decrease systematically with increasing ion fluence, revealing irradiation-induced changes that are hidden in standard dark measurements. Atomic force microscopy and X-ray photoelectron spectroscopy show substantial carbon-containing residues on processed devices even after extended cleaning. We propose that such residues may provide a reservoir for hydrocarbon-mediated passivation of sulfur vacancies. Density-functional-theory calculations provide a microscopic model consistent with this scenario: unsaturated V\textsubscript{S} introduce in-gap states, H--C$_\mathrm{S}$ configurations suppress these states, and carbon substitution without hydrogen leaves defect states in the band gap. Our results highlight carbon-containing surface contamination as a key factor in interpreting defect engineering experiments on \ce{MoS2} and related TMDC devices.
\end{abstract}

% keywords can be removed

\section{Introduction}\label{sec1}
Defect engineering is a central strategy for tailoring the electronic and optoelectronic properties of two-dimensional transition-metal dichalcogenides (TMDCs) \cite{Lu.2018, Pelella.2020, Sleziona.2023, Bertolazzi.2017, Jadwiszczak.2019}. In \ce{MoS2}, sulfur vacancies are among the most relevant point defects because they can introduce in-gap states, modify the carrier density, and act as scattering or trapping centers \cite{Bertolazzi.2017, Zhang.2024}. Such vacancies may arise during growth, processing, plasma exposure, electron irradiation, or ion irradiation. However, the electronic signature of these defects in processed field-effect transistors (FETs) is not necessarily straightforward. Contacts, adsorbates, substrate-induced charge trapping, and fabrication residues can strongly affect the measured transfer characteristics \cite{Pelella.2020, DiBartolomeo.2017}. In addition, chemically active vacancy sites may react with residual species from the sample surface or environment, thereby masking their expected electronic impact \cite{Fekri.2024}.

Monolayer \ce{MoS2} is a key model system for studying this problem because it combines a direct optical band gap with compatibility with field-effect transistor geometries \cite{Mak.2010, Radisavljevic.2011}. Its high surface-to-volume ratio and pronounced optoelectronic response make its transport properties sensitive to light, adsorbates, and the dielectric environment \cite{LopezSanchez.2013, DiBartolomeo.2017, DiBartolomeo.2018}. Device parameters such as doping, memory behavior, carrier mobility, and contact resistance can therefore vary strongly between processed devices and must be considered when interpreting defect-induced changes. Previous studies have shown that \ce{MoS2}-based devices can be modified by electron irradiation, ion irradiation, chemical treatments, and surface adsorbates \cite{Lu.2018, Pelella.2020, Sleziona.2023, Bertolazzi.2017, Jadwiszczak.2019}, but the resulting device response often reflects a combination of intrinsic defects, extrinsic residues, and interface effects.

Several reports indicate that \ce{MoS2} FETs can display an unexpectedly weak or non-monotonic degradation of their dark electrical characteristics after ion irradiation \cite{Fekri.2024, Fox.2015}. Such apparent robustness may originate from substrate-related effects, contact modifications, or chemical passivation of irradiation-induced defects. In particular, it has been suggested that chemically active sulfur vacancies may react with sputtered or residual species, thereby masking the direct impact of irradiation on the channel \cite{Fekri.2024}. More generally, this raises the question whether defect-rich \ce{MoS2} devices necessarily reveal their defect density in standard dark electrical measurements, or whether chemically passivated defects can remain electronically hidden.

In this work, we use low-energy Ar$^+$ irradiation as a controlled route to introduce sulfur vacancies into \ce{MoS2} FETs. The ion energy is chosen to preferentially create near-surface defects in the \ce{MoS2} channel while minimizing damage to the substrate and contacts. The dark transfer characteristics suggest an apparent robustness against irradiation up to moderate fluences, with pronounced degradation only at the highest fluence. In contrast, measurements under 532~nm illumination reveal a systematic reduction of the photoresponse with increasing ion fluence. This discrepancy shows that the defect response is conditional and can be overlooked when only dark electrical characteristics are considered.

We examine whether this apparent defect saturation can be explained by carbon-containing surface residues. Carbon contamination is common in processed and synthetic TMDC samples and may originate from growth precursors, ambient exposure, or lithographic processing \cite{Cochrane.2021, cochrane2020intentional, schuler2019substitutional, park2020effect, zhang2019carbon, Park.2023, Liu.2026}. In our processed \ce{MoS2} devices, surface analysis reveals substantial carbon-containing residues, including hydrocarbon-like components. Such residues provide a plausible reservoir of carbon-containing fragments that may interact with sulfur vacancies generated during ion irradiation.

This scenario is motivated by recent atomic-scale studies of carbon-related defects in TMDCs. In \ce{WS2} and \ce{WSe2}, carbon-hydrogen (CH) impurities have been identified as persistent unintentional defects, and CH incorporation at chalcogen sites has been shown to be energetically favorable \cite{Cochrane.2021, cochrane2020intentional, schuler2019substitutional}. In \ce{WS2}, a CH complex at a chalcogen vacancy yields a binding energy of $-0.55$~eV \cite{cochrane2020intentional}. Moreover, CH produces less lattice strain than C, \ce{CH2}, or \ce{CH3} configurations \cite{zhang2019carbon}. Although these results were obtained mainly for \ce{WS2} and \ce{WSe2}, they establish CH substitution at chalcogen sites as a chemically plausible motif in TMDCs.

Based on these considerations, we propose hydrocarbon-mediated passivation of sulfur vacancies as a plausible microscopic scenario. In particular, H-C$_\mathrm{S}$ configurations provide a model system for understanding how vacancy-related in-gap states could be suppressed, thereby masking the effect of defects in dark transfer measurements. The observed degradation under illumination then suggests that the chemical or electronic state of these passivated defects may be modified by light. We support this interpretation by combining dark and illuminated electrical measurements with surface analysis and density-functional-theory calculations.

\section{Results and Discussion}
%Ich habe hier hauptsächlich die reihenfolge geändert (immer erst bedingungen, dann ergebnis, dann interpretation, nicht wild durcheinander), gestrafft und sprachlich geglättet
\begin{figure}
    \centering
    \includegraphics[width=0.75\linewidth]{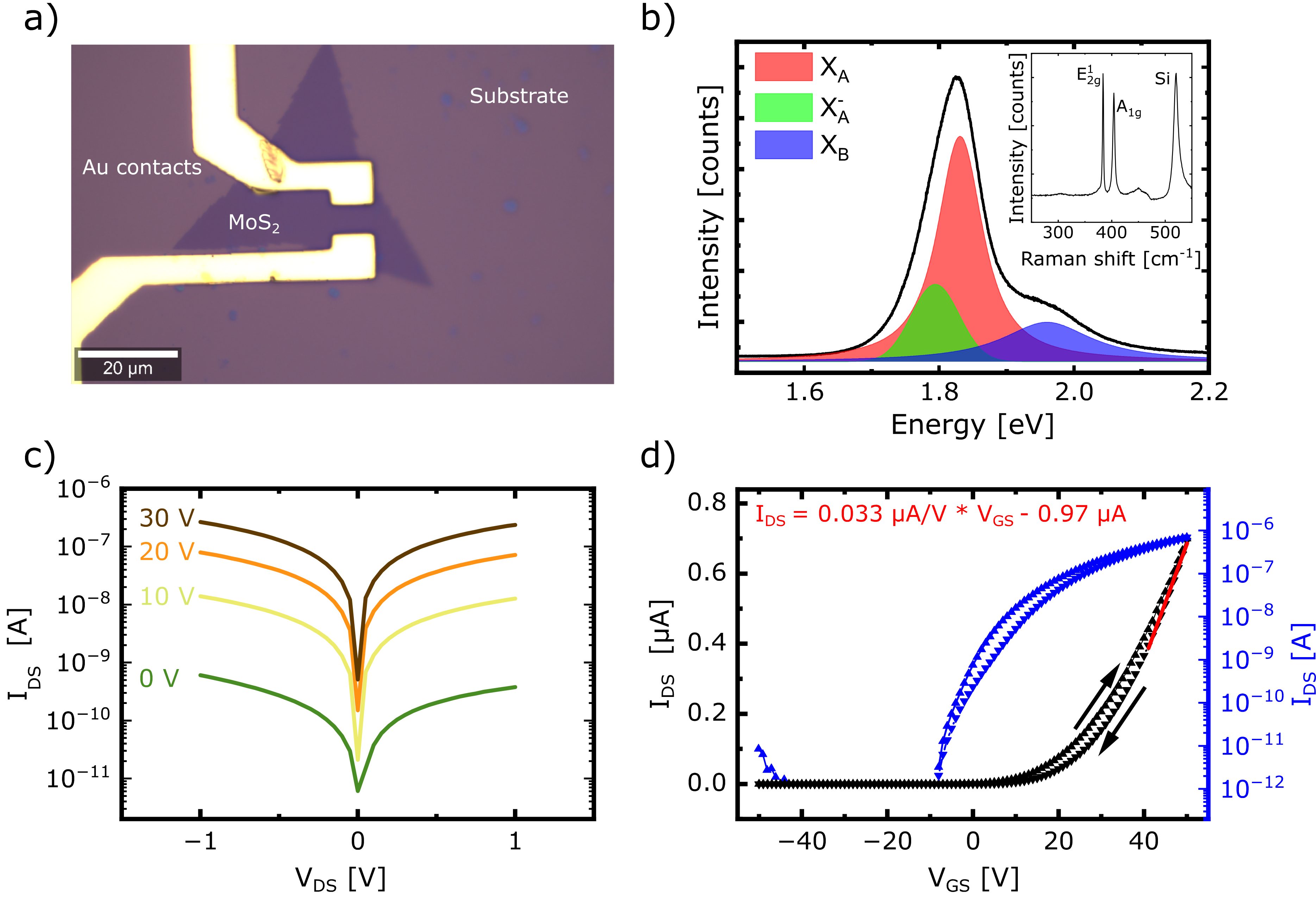}
    \caption{(a) Microscopic image of an \ce{MoS2} monolayer with Cr/Au contacts on a \ce{SiO2} substrate. (b) Pl spectrum of \ce{MoS2} confirming the monolayer properties of the flake. The insert shows the corresponding Raman spectra. (c) Output characteristics of an \ce{MoS2} FET. The output current for negative and positive V\textsubscript{DS applied} is rather symmetric, with a ratio of $\approx1.1$, indicating mostly ohmic behavior. (d) Transfer characteristics of an \ce{MoS2} FET with linear and logarithmic y-axis. The backswept measurement is linearly fitted for a range from 40~V to 50~V for extracting the threshold voltage and carrier mobility.}
    \label{fig1}
\end{figure}

We begin by establishing the baseline properties of the \ce{MoS2} FET prior to ion irradiation. An optical micrograph of the device is shown in Figure~\ref{fig1}(a). The monolayer \ce{MoS2} was grown by chemical vapor deposition (CVD) directly on p-doped \ce{SiO2}/Si wafers.

As a first characterization step, we confirm the monolayer character of the \ce{MoS2} flake by photoluminescence and Raman spectroscopy, as shown in Figure~\ref{fig1}(b). The photoluminescence spectrum exhibits the characteristic emission features of monolayer \ce{MoS2}, associated with excitonic recombination processes  \cite{Mak.2010, Scheuschner.2014, Pollmann.2020}. In addition, the A exciton peak shows asymmetric properties with a low-energy tail, consistent with the presence of trions and thus with electron doping of the monolayer. \cite{Christopher.2017}. The Raman signal is consistent with as-grown monolayer \ce{MoS2}, with the A$_\text{1g}$ and E$^1_\text{2g}$ modes separated by approximately 20~cm$^{-1}$\cite{Pollmann.2020}.

%Bisschen wie ein Laborbericht  
The as-grown flake was subsequently processed into a FET by defining \ce{Cr/Au} source and drain contacts using standard photolithography. The Cr layer serves as the adhesion layer and forms the direct contact to \ce{MoS2}, while Au provides the highly conductive top electrode. Cr has been proposed as a favorable contact metal for \ce{MoS2} because of its suitable work function and high density of states near the Fermi level \cite{Luo.2015, Yang.2021}. Nevertheless, the contact properties of processed \ce{MoS2} FETs can be strongly affected by interface contamination, oxidation, or lithographic residues, which may introduce additional barriers and increase the contact resistance \cite{Yang.2021}. The degenerately doped Si substrate was used as a global back gate.

Lithographic processing can leave residual organic material on the \ce{MoS2} surface. To assess this contribution, we investigated a processed \ce{MoS2} reference sample by atomic force microscopy (AFM). After a standard lift-off process, the surface remains covered by nanoscale residues, visible as protrusions in the AFM topography. Repeated acetone cleaning ($3\times$4~h) substantially reduces the surface roughness from 1.9~nm to 0.3~nm. However, residual material can still accumulate near the Au contacts, indicating that the fabrication process does not fully remove surface contamination. As discussed below, XPS confirms that these residues contain substantial carbon.

%Device under vacuum
%Cr/Au as spurce and drain
%Fig1c
%nearly symmetric / no dominant rectification
Electrical measurements are done for further characterization of the sample. For all electrical measurements, the device is placed under high vacuum ($\leq5\cdot10^{-5}$~mbar) to minimize the influence of adsorbates and influence of the atmosphere. The \ce{Cr/Au} contacts act as drain and source contacts, enabling the performance of output measurements, where the drain-source current (I\textsubscript{DS}) is measured in dependence of the drain-source voltage (V\textsubscript{DS}). The output characteristic of the sample is shown in Figure~\ref{fig1}(c). V\textsubscript{DS} is swept from -1~V to 1~V and back, with the measurement sweep backwards being displayed. The measurements were repeated for different voltages, measured relative to the potential at the source contact, applied at the gate (V\textsubscript{GS}). The nearly symmetric output characteristics provide little evidence for a dominant rectifying Schottky barrier and indicate predominantly ohmic behavior in the measured voltage range\cite{Grillo.2021,Schulman.2018,Pelella.2020}. 

%Absatz2 Transfer measurment
%V_DS=0.5V
%Sweep protocol
%Fig 1d
%n-type
%&ON/OFF
%residual CH
For the transfer measurements, V\textsubscript{DS} is held constant at 0.5~V with the V\textsubscript{GS} twice swept from -50~V to 50~V back to -50~V in 1~V steps. After the gate voltage is changed, a delay time of 3.6~s is kept, before the output signal by the picoammeter is integrated for 25~ms. These gate-dependent measurements, shown in Figure~\ref{fig1}(d), indicate pronounced n-type doping, as commonly observed for \ce{MoS2} FETs\cite{LopezSanchez.2013, Desai.2016, DiBartolomeo.2017, DiBartolomeo.2018, Sleziona.2023, Bertolazzi.2017}. The device shows clear on-off behavior above $10^3$, typical for our devices \cite{Sleziona.2023}. A possible contribution to this n-type behaviour is carbon incorporation during growth  \cite{Park.2023}, potentially through the precursor used in CVD-growth. Other descriptors introduced in this work, such as carrier mobility $\mu$ or doping $n$, refer to electrons as the majority charge carriers. 

%Absatz3
%Hysterese
%Typisch
%im vakuum oxide trapping
%SI
The device displays a typical hysteresis effect commonly observed in \ce{MoS2} FETs \cite{DiBartolomeo.2018, Kumar.2023}.  
Different mechanisms are discussed as the origin of the hysteresis. Fast recombination through intrinsic in-gap states is unlikely to explain the slow hysteresis observed here, as such processes typically occur on ultrafast timescales ($\leq$10~ns) \cite{Zheng.2026, Li.2019, Yuan.2015}.It was shown that the atmospheric environment can impact the hysteresis \cite{AlMamun.2024}. As our device operates under vacuum conditions, charge trapping at the interface and in the oxide seems to be the most likely mechanism \cite{DiBartolomeo.2018, Farronato.2023, Illarionov.2020, Illarionov.2017, Stampfer.2018}, with the gate voltage pushing electrons in and out of these states over time. This process happens in the order of several minutes, which  also holds true in our device, as can be seen in time-resolved gate-dependent measurements (See SI).

%Absatz4 Mobility
%effective field mobility
%formel
%value
%interpretation
The effective mobility is determined through linear fitting the transfer curve for a fixed range at high voltages (Figure~\ref{fig1}(d)) through the following expression:
\begin{equation}
    \mu = \frac{\delta I_\text{DS}}{\delta V_\text{GS}}\frac{1}{C_\text{ox}V_\text{DS}}\frac{L}{W}
\end{equation}

$W$ and $L$ are the width and length of the conductive channel, with $C_\text{ox} = 1.21\times10^{-8}$~F/cm$^2$ being the capacitance per unit area of the oxide \cite{Sleziona.2023}. We estimate the mobility of the device to be around 4.3~cm$^2$/V$\cdot$s, typical for such devices \cite{DiBartolomeo.2017, DiBartolomeo.2018, Sleziona.2023, Zafar.2017, Mu.2025, Zhu.2014}. 
The effective mobility $\mu$ in the channel material is connected with the mobility of free electrons $\mu_0$ in the conduction band, the overall electron concentration $n$, and the concentration of electrons in extended states $n_\text{band}$ with $\mu= \mu_0\frac{\delta n_\text{band}}{\delta n}$ \cite{Mu.2024, Yu.2017}. Electrons in extended states refer to electrons that contribute to band transport. Carrier trapping results in a reduction of the effective mobility. 

%Absatz5 limits/scattering
%typisch
%phonon limit: matthiesse habe ich raus genommen, da wir die einzelnen channel ja eh nich ansprechen
%reale devices
%hier vermutlich auch
The extracted mobility is well below the room-temperature phonon-limited value of approximately 150~cm$^2$/V$\cdot$s reported for \ce{MoS2} \cite{Mu.2024, Yu.2016, Li.2013}. This indicates that additional scattering and trapping mechanisms contribute substantially in the processed device \cite{Zhu.2014, Mu.2024, Mu.2025, Yu.2017}. %At room temperature the mobility of \ce{MoS2} devices is limited to around 150~cm$^2$/V$\cdot$s due to phonon scattering \cite{Mu.2024, Yu.2016, Li.2013}. Surface impurities or charged defect enable Coulomb scattering. In CVD-grown samples structural defects and carrier trapping is identified to dominate and limit mobility \cite{Zhu.2014, Mu.2024, Mu.2025, Yu.2017}. We suggest that the same holds true for our devices. 

\begin{figure}
    \centering
    \includegraphics[width=1\linewidth]{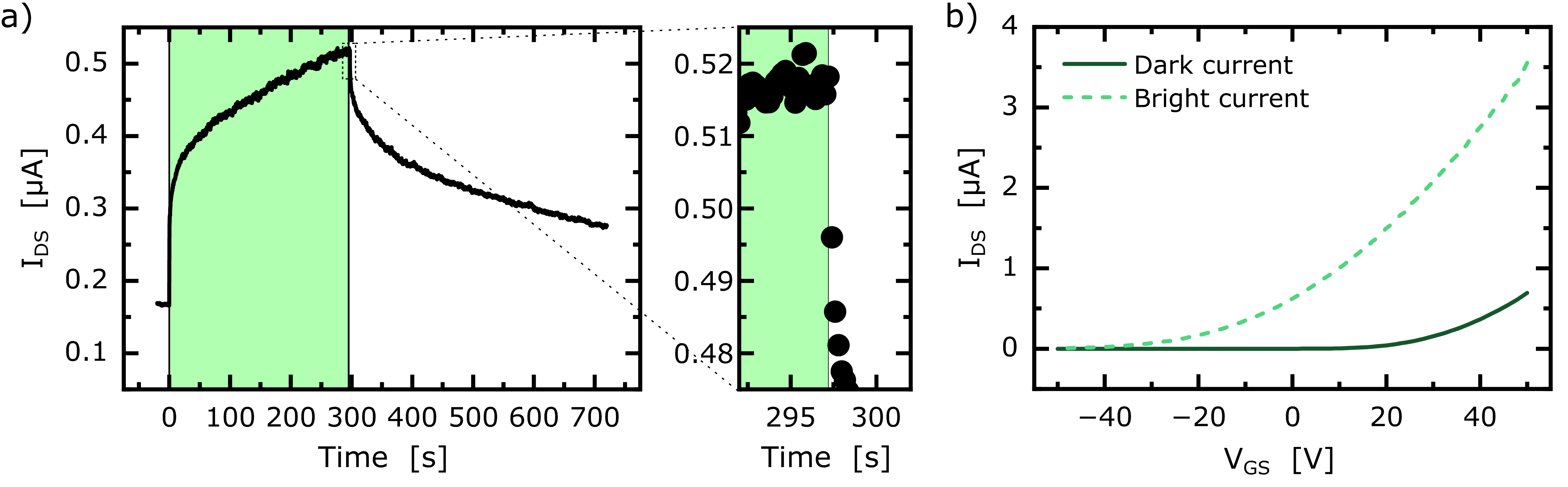}
    \caption{(a) Time-resolved photocurrent measurements for \ce{MoS2}. The channel is illuminated for around 300~s with a 532~nm laser. The duration of the illumination is marked in the figure. A zoomed-in look at the data shows that the current changes only minimally immediately (200~ms) after the laser is turned off compared to the total current induced by the laser, indicating that direct photoexcitation is not the dominant contribution to the persistent photocurrent on the observed timescale. (b) Transfer characteristic for an \ce{MoS2} FET under dark and under bright conditions through laser illumination. A shift of the transfer curve to the left indicates how the laser effectively n-dopes the material.}
    \label{fig2}
\end{figure}

The optoelectronic response of the device was characterized by illuminating the \ce{MoS2} channel with a 532~nm cw laser at a total power of 0.1~mW. As shown in Figure~\ref{fig2}(a), illumination leads to a pronounced increase in the drain-source current. The current rises gradually after an initial jump and does not immediately return to its original value after the light is switched off, which is typical for persistent photoconductivity in \ce{MoS2} FETs \cite{DiBartolomeo.2017, Furchi.2014}. The transient response requires at least two characteristic timescales to describe, with $\tau_1\approx15$~s and a slower component $\tau_2$ exceeding the observed time window \cite{DiBartolomeo.2017}.

%figure gehört in diesen Abschnitt

Different processes contribute to the photocurrent, including the direct photoexcitation, desorption of p-doping adsorbates and trapping of carriers \cite{DiBartolomeo.2017}. Due to the short lifetime of photogenerated carriers, direct photoexcitation is expected to contribute mainly to the fast component of the photocurrent and is unlikely to dominate the persistent response observed here \cite{DiBartolomeo.2017,Li.2019,Zheng.2026}. This can be seen in Figure~\ref{fig2}(a), where the current changes only slightly (20~nA) 200~ms after the laser is turned off, compared to the overall current induced by the laser in the presented measurement. 
We therefore attribute the persistent photocurrent primarily to slow processes such as desorption of p-doping adsorbates and charge trapping. In Figure~\ref{fig2}(b) the effective doping through the laser can be observed through a shift of the transfer curve to the left. For a better overview, we only plot the current for the measurements swept backwards.

The photodoping density is estimated from the light-induced shift of the threshold voltage $V_\text{th}$. For both dark and illuminated transfer curves, $V_\text{th}$ is obtained from the intersection of a linear fit with the gate-voltage axis, as illustrated in Figure~\ref{fig1}(d).The corresponding photodoping density is calculated as $n_\text{ph} = C_\text{ox}\times \frac{V_\text{th}(\mathrm{Dark})-V_\text{th}(\mathrm{Bright})}{e}$ \cite{Gadelha.2020}. For the measurements shown, we see a strong increase in n-doping of about $1.8\times10^{12}$~1/cm$^2$ induced by the laser.
%\textcolor{red}{Stimmt das mit dem Vorzeichen? Der Kurvenshift sagt ja was anderes - das hatte ich doch schon angemerkt im pdf?. Müsste es nicht eher sein:} $n_\text{ph} = \frac{C_\text{ox}}{e}\left[V_\text{th}(\mathrm{Dark}) - V_\text{th}(\mathrm{Bright})\right]$

%Warum 40 ev Ar
%Experiment
%Dark transfer / Auswertung Fig3a
%Überraschung
%Fekri: Oxidtraps?
%Memory window
Low-energy Ar$^+$ irradiation is used to introduce defects into the \ce{MoS2} channel. The ion energy is set to 40~eV, where near-surface defect creation in \ce{MoS2} is favored while substrate sputtering and the formation of deep oxide defects are expected to be strongly reduced \cite{Kretschmer.2018, Ziegler.2010}. In this regime, sulfur vacancies are expected to be the dominant irradiation-induced defects, with a high probability of vacancy creation per incident Ar$^+$ ion \cite{GhorbaniAsl.2017}. 

\begin{figure}
    \centering
    \includegraphics[width=1\linewidth]{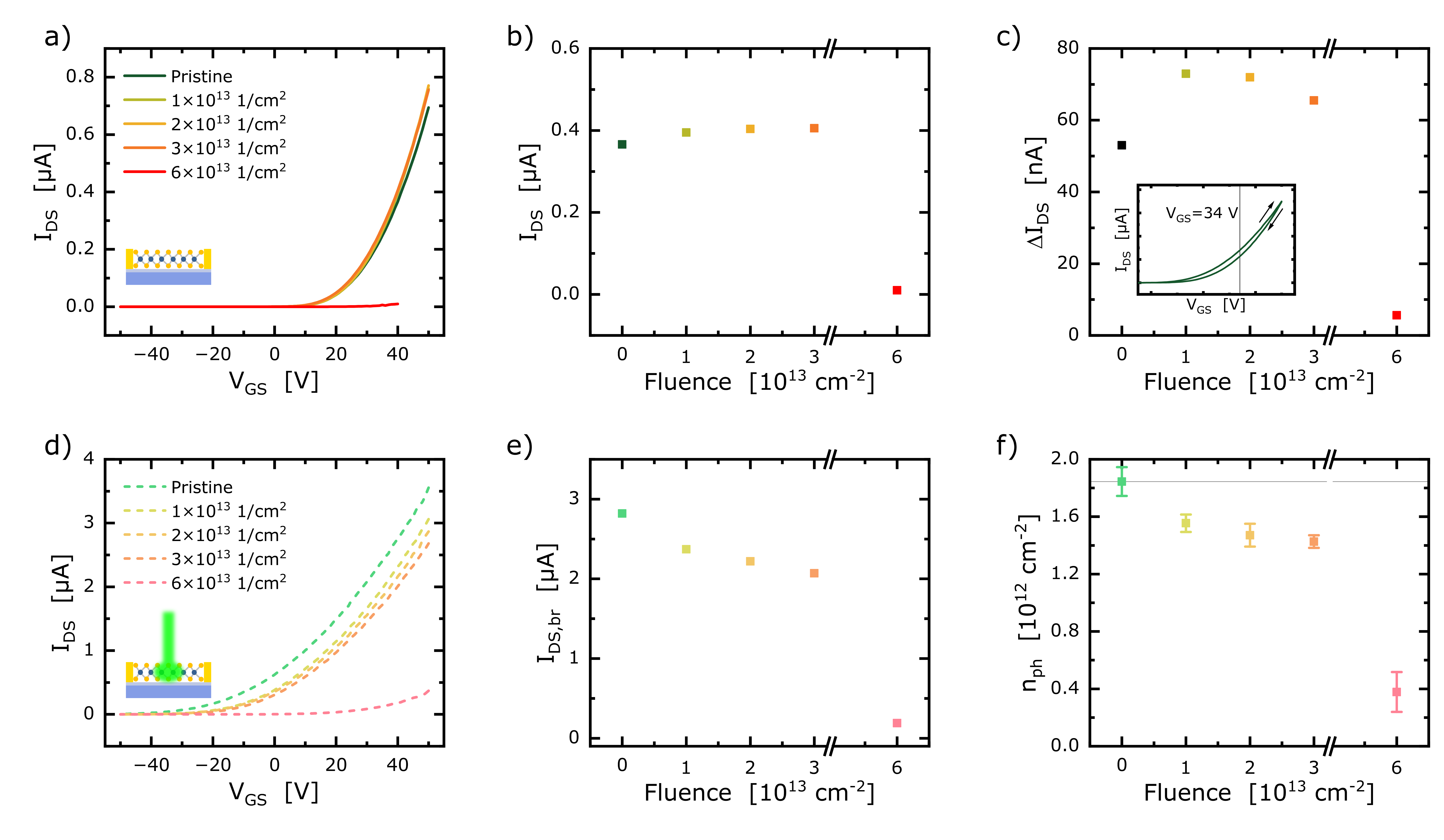}
    \caption{(a) Transfer curve for an \ce{MoS2} FET under dark conditions for different fluences of Ar$^+$ ions. Through fitting the transfer curve at high voltages, the mobility and threshold voltage are extracted. (b) $I_\text{DS}$ extracted from (a) at V$_\mathrm{GS}$=40 V. The current remains nearly unchanged up to a fluence of 3$\times 10^{13}$ions/cm$^2$ and only decreases for high fluences. (c) Memory window extracted from the hysteresis for each irradiation step. The memory window was measured at a gate voltage of V$_\mathrm{GS}$=34 V as indicated in the insert, where the memory window is the widest for this device. The limited changes suggest that ion-induced oxide trapping is not the dominant contribution under these irradiation conditions. (d) Transfer curve for an \ce{MoS2} FET under 532~nm illumination for different irradiation steps. (e) $I_\text{DS}$ extracted from (d) at a gate voltage of 40~V. In contrast to the measurements under dark conditions, the current decreases with each irradiation step. (f) Photo doping, additional carriers induced by the laser, for each irradiation step.}
    \label{fig3}
\end{figure}

For irradiation the sample is taken out of the analysis chamber and placed into an ultra-high vacuum (UHV) chamber equipped with an ion sputter gun. During this transfer process the sample is exposed to ambient conditions for a short period of about 5~minutes. Before conducting the ion irradiation, the sample is left to rest under UHV conditions ($10^{-8}$~mbar) for at least 12 hours to reduce weakly bound atmospheric adsorbates. After each irradiation step, the sample is taken out of the UHV and placed back into the analysis chamber, where the sample is again characterized. The corresponding transfer curves under dark conditions are shown in Figure~\ref{fig3}(a).

For each fluence, the mobility is determined. Additionally, changes in doping relative to the initial device state before irradiation are determined similarly as the photodoping with:
\begin{equation}
    \Delta n_\text{D} = C_\text{ox}\times \frac{\Delta V_\text{th}}{e}
\end{equation}

We note that charge accumulation on the oxide surface can lead to oxide breakdown at higher irradiation fluences. In the device shown here, a significant gate-source leakage current appears after an irradiation fluence of $6\times10^{13}$~cm$^{-2}$ for gate voltages above 10~V. The measured drain-source current was therefore corrected by subtracting the measured leakage current.

If unsaturated sulfur vacancies dominated the electronic response, one would expect a decrease in current and mobility already at low to moderate fluences. Instead, the current remains nearly unchanged during the first three irradiation steps (Figure~\ref{fig3}(b)). This is also seen in the mobility and doping (See SI). Only at a high fluence a strong decrease in the current can be observed. This behavior was observed in two reference samples, where the transfer characteristic stays constant with several irradiation steps only to significantly reduce for high fluences (see Supporting Information and \cite{Krasheninnikov.2026}).

Fekri \textit{et al.} reported a related non-monotonic response in \ce{MoS2} FETs irradiated with 5~keV and 7.5~keV He$^+$ as well as 7.5~keV Ne$^+$ ions. They attributed the increased conductivity at low and intermediate fluences to charges implanted in the oxide, which effectively act as an additional gate \cite{Fekri.2024}.
Only at higher fluences does the ion irradiation lead to the current being reduced. 

This mechanism would be expected to modify the hysteresis, since long-lived oxide or interface traps contribute to the memory window of \ce{MoS2} FETs. We therefore extract the memory window from the forward and backward gate sweeps for each irradiation step (Figure~\ref{fig3}(c)). Defects created in the oxide increase the hysteresis by acting as long-time charge traps \cite{Sleziona.2025}. We observe limited change in the memory properties of the used \ce{MoS2} FET. After the first irradiation step an increase in the hysteresis width of around 25\% can be seen, which the remains constant for the next two irradiation steps and decreases for higher irradiation steps. In a previous study, we irradiated \ce{MoS2} with fast (180~keV) highly charged Xe$^{30+}$ ions, depositing energy deep in the oxide, led to a consistent increase in the memory window in contrast to the measurements presented in this work \cite{Sleziona.2023}. This indicates that the irradiation does not lead to a pronounced increase in long-lived oxide or interface traps that would dominate the hysteresis response.

%Bright reveals hidden effect
%unstaurated V_S cannot explain this
%O passivation possible but not dominant
%CH passivation hypothesis
%H-C_S plausible model
%--> DFT
The transfer measurements were then repeated under illumination with a 532~nm laser, as shown in Figure~\ref{fig3}(d).. Compared with the dark measurements, the illuminated transfer curves show a markedly different fluence dependence. Already at low fluences the drain-source current decreases under illumination (Figure~\ref{fig3}(e)), accompanied by a significant reduction in photo-doping (Figure~\ref{fig3}(f)). 

The measurements under illumination and under dark conditions show seemingly contradictory results. The comparison between dark and illuminated measurements suggests that irradiation-induced changes can be hidden under dark measurement conditions.

If the irradiation-induced sulfur vacancies remained electronically active and unsaturated, they would be expected to introduce localized in-gap states. These states do not necessarily act as shallow reversible traps, but they can still affect transport by localizing charge, modifying the apparent carrier density, promoting recombination under illumination, or increasing charged-defect scattering \cite{Bertolazzi.2017}. The weak changes observed in the dark transfer characteristics are therefore difficult to reconcile with a large population of electronically active, unsaturated sulfur vacancies alone.

A plausible explanation for this apparent masking is an initial chemical passivation of the sulfur vacancy sites. One possible passivation pathway is oxygen incorporation into sulfur vacancies, with oxygen supplied either by substrate sputtering \cite{Fekri.2024} or by atmospheric exposure during sample transfer. Oxygen is known to occupy V\textsubscript{S} sites in TMDCs and can form relatively inert substitutional defects \cite{Daniel.2025, Frammolino.2025}. Once atomic oxygen is incorporated into a sulfur vacancy, its removal is expected to be energetically unfavorable, as the formation energy to remove substitutional oxygen exceeds that of sulfur removal from the \ce{MoS2} lattice \cite{Wu.2020, Zhang.2024}. Oxygen substitution of irradiation-induced vacancies is therefore unlikely to fully account for the observed behavior.

This argument is strengthened when we consider two possible oxygen sources. First, SRIM calculations indicate that, at the ion energy used here, substrate sputtering is negligible. Substrate-derived oxygen species are therefore unlikely to provide a dominant source for oxygen passivation. This is in accordance with the limited change in the memory window, which suggests that long-lived oxide or interface traps do not dominate the irradiation response. Secondly, passivation of V\textsubscript{S} sites by incorporating atmospheric oxygen is expected to be kinetically limited under ambient conditions, with reported timescales on the order of days due to the activation barrier for chemisorption \cite{Luo.2022}. We therefore consider oxygen-based passivation unlikely to dominate on the relevant experimental timescale.

\begin{figure}
    \centering
    \includegraphics[width=0.35\linewidth]{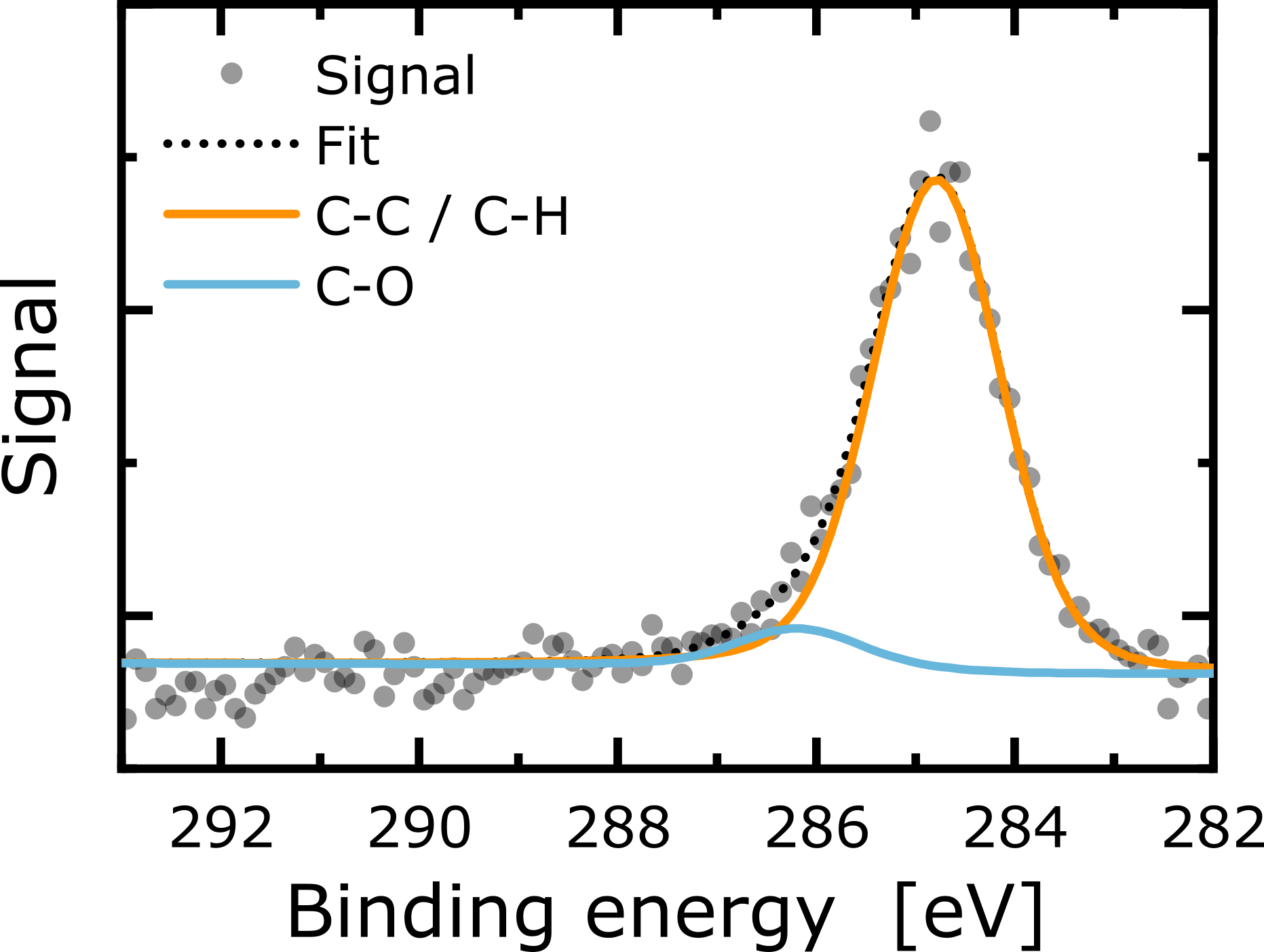}
    \caption{XPS spectrum of the C~1s region measured on a processed \ce{MoS2} reference sample after cleaning. The spectrum is consistent with hydrocarbon-like residues and partially oxidized carbon species. C-C and C-H bonds contribute to $\approx$90\%  and C-O bond $\approx$10\% to the signal.}
    \label{fig:XPS}
\end{figure}

Alternatively, we propose that carbon-related defect formation should be considered. Recent studies have shown that TMDCs frequently exhibit carbon contamination \cite{Cochrane.2021, cochrane2020intentional, schuler2019substitutional, park2020effect, zhang2019carbon, Park.2023, Liu.2026}. Carbon-containing precursors, such as sodium cholate used in this study, are commonly used during growth. Subsequent device fabrication also exposes the material to additional organic species, i.e. photoresist material. 
X-ray photoelectron spectroscopy (XPS) analysis of a processed \ce{MoS2} reference sample after cleaning reveals substantial carbon-containing surface contamination (Figure~\ref{fig:XPS} and SI). The C~1s spectrum contains a dominant low-binding-energy carbon component, contributing to $\approx$90\% of the carbon-related signal, commonly assigned to C-C/C-H environments, together with a higher-binding-energy C-O contribution, contributing to $\approx$10\% of the carbon-related signal. This is consistent with hydrocarbon-like residues and partially oxidized organic contamination.

%X-ray photoelectron spectroscopy (XPS) analysis of the processed \ce{MoS2} FET device (Figure~\ref{fig:XPS}) reveals substantial amount of carbon contamination as the most prevalent surface impurity. 

In light of this, previous studies have also reported alkane-covered TMDC surfaces under ambient conditions \cite{Palinkas.2022}. Therefore, we argue that such surface contamination represents a readily available reservoir of carbon-containing species, whose role in irradiated TMDC-based FETs may be overlooked.

%the y-axis on this spectra is better off as "intensity (a.u.)
These contaminants may interact with sulfur vacancies generated during ion irradiation. In addition, ion bombardment can fragment surface hydrocarbons and thereby create reactive hydrocarbon fragments and radicals \cite{Cheng.2023}. We therefore hypothesize that ion-induced sulfur vacancies is highly susceptible to passivation by carbon-containing fragments derived primarily from residual surface organics. Additional contributions from hydrocarbon background species in the vacuum system or from ambient exposure during sample transfer cannot be excluded. 

The resulting carbon-related vacancy complexes are expected to differ from substitutional oxygen defects. Oxygen incorporation can fully saturate the sulfur vacancy and form a comparatively stable O$_\mathrm{S}$ configuration. Carbon incorporation, by contrast, may leave the defect site chemically or electronically unsaturated unless additional moiety are involved. In related TMDC systems, this H-C\textsubscript{S} configuration has been identified as energetically favorable carbon-related defects \cite{Cochrane.2021, cochrane2020intentional, schuler2019substitutional, park2020effect, zhang2019carbon}. Deliberate introduction of H-C\textsubscript{S} in \ce{WS2} FETs produced transport trends that are qualitatively similar  to our dark measurements, including an initial increase in current followed by degradation at higher concentrations \cite{zhang2019carbon}. Although the host material and doping response differ, these results demonstrate that H-C\textsubscript{S} defects can substantially modify the transport properties of TMDC FETs. This makes H-C$_\mathrm{S}$ a plausible microscopic model for carbon-mediated vacancy passivation in the present \ce{MoS2} devices.

Accordingly, we suggest that carbon-mediated passivation can suppress vacancy-related in-gap states under dark conditions, while illumination may modify the electronic or chemical state of these passivated defects, and thereby reveal the irradiation-induced changes in the photoresponse.

\begin{figure}
    \centering
    \includegraphics[width=0.8\linewidth]{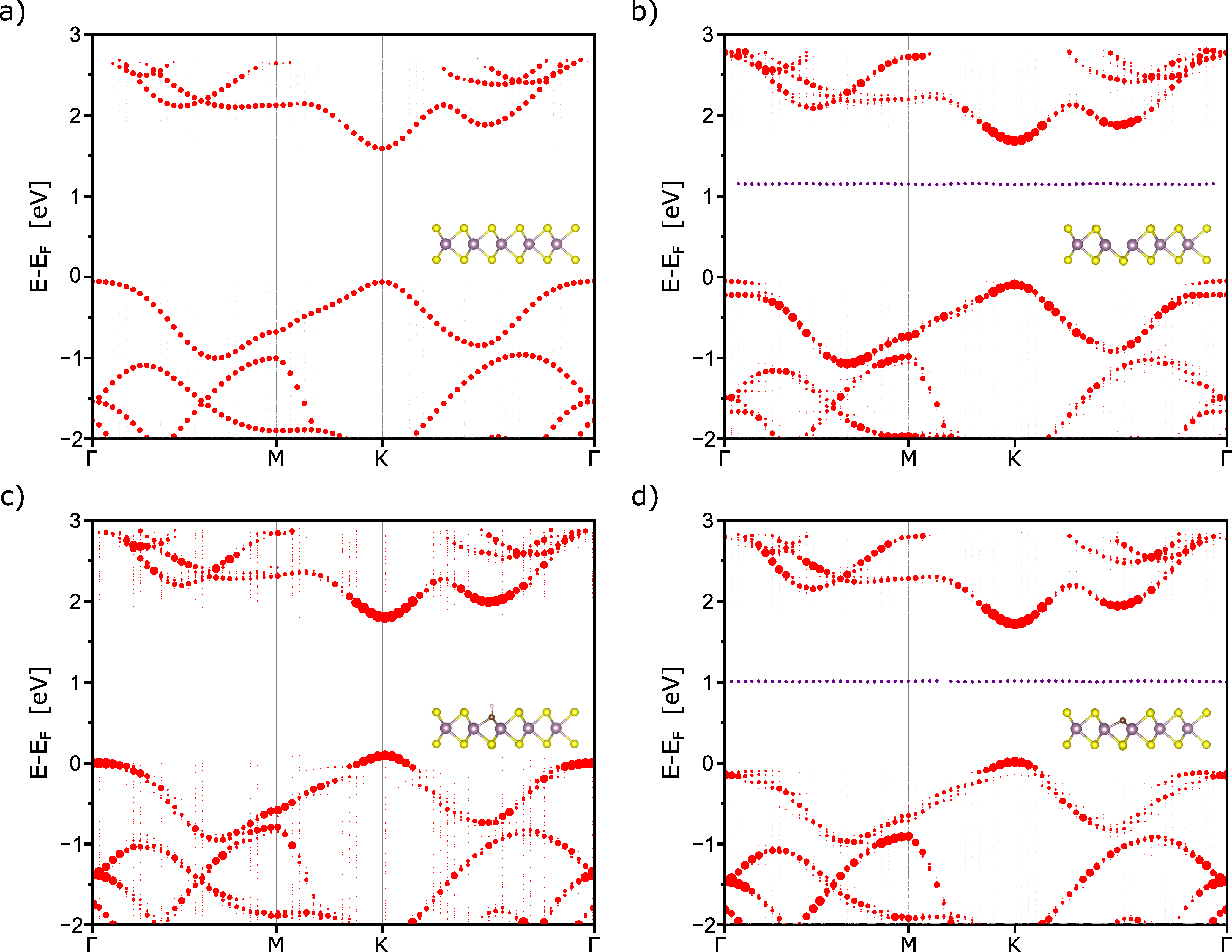}
    \caption{Unfolded band structures from monolayer \ce{MoS2} $5 \times 5$ supercell, pristine (a), sulfur vacancy (V\textsubscript{S}) (b), CH substitution (H-C\textsubscript{S}) on sulfur vacancy (c),  carbon occupying sulfur vacancy (C\textsubscript{S}) (d). Typical in-gap states stemming from V\textsubscript{S} are saturated with CH substitution.}
    \label{fig4}
\end{figure}

To further corroborate our hypothesis, we have modeled the occupation of a V\textsubscript{S} site by C\textsubscript{S} and H-C\textsubscript{S} via DFT calculations (For details see Method section). In Figure~\ref{fig4}(a)-(d) we show the band structure of pristine \ce{MoS2}, with a sulfur vacancy, and the described modification. The presence of a sulfur vacancy induces three gap states (Figure~\ref{fig4}(b)) coming predominantly from the Mo-d orbital: one occupied bonding level, located just above the valence band, and a pair of energy-equivalent, unoccupied antibonding levels (purple). The H-C\textsubscript{S} addition effectively saturates the dangling bonds of Mo (Figure~\ref{fig4}(c)), removing mid-gap states that can act as trapping centers. As no new carrier traps are introduced, this provides a plausible microscopic model for the absence of strong dark-current degradation in our measurements.
%we observe no initial negative impact on the current in our measurements. 

To assess the thermodynamic stability of the defect configurations, we calculated their respective formation energies. The formation energy ($E_f$) for a pristine sulfur vacancy ($V_{\text{S}}$) was determined to be $3.01$~eV. In comparison, the formation of $\text{H-C}_{\text{S}}$ yielded an $E_f$ of $2.76$~eV. Using the $V_{\text{S}}$ configuration as a reference, the incorporation of the $\text{H-C}$ group into the vacancy site is exothermic by $-0.25$~eV, indicating that the plugging process is energetically favorable.  

Band structure calculation on Figure~\ref{fig4}(d) shows that carbon occupation alone fails to fully saturate V\textsubscript{S} sites, leaving behind an in-gap state. Defect states reduce the population of mobile carriers in the conduction band while increasing the number of trapped carriers that do not participate in charge transport, thereby lowering the effective mobility \cite{Bertolazzi.2017, Mu.2024, Yu.2017}. 
%PK 
%To investigate the effect of heating by the laser
%MS: Aber das ist doch nicht der hintergund, weil wir doch nicht über die mögliche erwärmung als ursache diskutieren, sondern inert vs. chemically/electronically labil, oder?
To evaluate whether H–C\textsubscript{S} should be regarded as an inert passivation configuration, we calculate the reaction enthalpy $\Delta H$ (at zero temperature and pressure) for the reaction $\text{H-C}_{\text{S}} \to \text{C}_{\text{S}} + \frac{1}{2}\text{H}_2 + \Delta H$ 
%The dissociation energy of hydrogen from  H-C\textsubscript{S} model was 
where $\Delta H$ was computed to be $-1.1$~eV. We suggest that H-C\textsubscript{S} formation at V\textsubscript{S} sites passivates vacancy-related gap states and thereby masks the effects of ion irradiation at low fluences in TMDC devices. The calculated reaction enthalpy indicates that dehydrogenated C\textsubscript{S}-like configurations are energetically accessible from H-C\textsubscript{S}, suggesting that these carbon-related defect configurations may be less inert than oxygen-passivated vacancies. Such configurational or electronic changes could contribute to the observed light-induced degradation of the photoresponse.
%MS
%We suggest that the substitution of V\textsubscript{S} sites with H-C\textsubscript{S} sites masks the effects of ion irradiation at low fluences in TMDC devices. However, these substitutions are, in contrast to oxygen substitutions, not inert and may easily react under illumination. 

\section{Conclusions}
%Dedi: Please check this section after you have finishe the DFT part
We show that processed \ce{MoS2} FETs can appear robust against low-energy ion irradiation when characterized under dark conditions. While the dark transfer characteristics remain nearly unchanged up to moderate fluences, photocurrent measurements reveal a systematic degradation with increasing ion fluence. This demonstrates that irradiation-induced changes can be hidden in standard electrical characterization and become apparent only under illumination.

We propose carbon-mediated passivation of sulfur vacancies as a plausible explanation for this behavior. XPS analysis reveals a significant carbon species contamination on processed \ce{MoS2} devices, providing a potential reservoir for hydrocarbon-derived fragments. 

Density functional theory (DFT) calculations yield a microscopic framework that aligns with this proposed mechanism: H-C$_\text{S}$ configurations suppress vacancy-related in-gap states, whereas carbon substitution without hydrogen leaves defect states in the band gap. The different behavior under dark and illuminated conditions suggests that illumination may modify the chemical or electronic state of such passivated defects, revealing irradiation-induced changes in the photoresponse. These findings underline the importance of carbon-containing surface contamination and light-dependent characterization for interpreting defect engineering in \ce{MoS2} and related TMDC devices.

\section{Methods}
\subsection{Sample Preparation}
Sample preparation followed a standard recipe \cite{Sleziona.2023}. For the CVD process, 2\% aqueous sodium cholate were spin-coated onto cleaned Si/\ce{SiO2}substrates as seeding promoters. Ammonium heptamolybdate (AHM, Sigma-Aldrich) was used as the Mo precursor. For converting AHM to \ce{MoO3}, 0.5~$\mu$L droplets were dispensed on one side of each substrate and annealed at $300\,^{\circ}\mathrm{C}$ for 24 min in air. The growth of \ce{MoS2} flakes was done in a three-zone hot-wall furnace with a horizontal cylindrical quartz reactor tube, with 500~sccm Argon serving as carrier gas. In the first zone the carrier gas was enriched with elemental sulfur (Sigma-Aldrich) at a temperature of $170\,^{\circ}\text{C}$. In the second zone, sulfur reacted with MoO$_3$ to form \ce{MoS2} monolayers at  $750\,^{\circ}\text{C}$. After the process, the furnace door was opened for cooling.

With an optical microscope, suitable flakes were selected for further processing. Through a standard photolithography process, 10 nm of Cr and 100 nm of Au contacts were deposited by electron-beam (Cr) and thermal evaporation (Au). The lift-off process is done by a 15~min long 80~$^\circ$C aceton bath, which includes the treatment with ultrasound for one minute, removing excess photoresist with Cr and Au. After the sample is rinsed with acetone, and ethanol.  

The samples are placed $3\times4$~h in acetone at 50~$^\circ$C bath while stirring constantly. The acetone is refreshed each time and rinsed with ethanol. All chemicals used are of analytic quality.

The dried substrates were then glued with silver conductive paint onto a customized sample plate and bonded. The sample plates enable the gating of the sample through a metal surface and simplify the connection for electrical measurements.

\subsection{PL and Raman Measurements}
PL and Raman measurements shown were conducted with a Witec Alpha300 RA setup using a 532 nm cw laser at 1~mW and a 50x (NA 0.55) objective from Zeiss. For PL measurements a grating of 300 lines/mm and for Raman measurements a grating of 1800 lines/mm was used.

\subsection{AFM}
All AFM measurements are done with the Witec Alpha300 RA setup in digital pulsed force mode. The resolution varies between 700 and 1090 pixels/$\mu$m$^2$. The setpoint was kept at 1.75~V and P gain and L gain at 4~\% and 3~\%. The driving frequency is 1000~Hz. For all measurements standard force modulation AFM probes
by NanoWorld were used.

\subsection{XPS}
XPS measurements were performed on a VersaProbe II by Ulvac-Phi using a monochromatic Al-$\alpha$, $\nu = 1486.7$~eV, beam, a beam diameter of 50~$\mu$m and an angle between sample surface and analyzer of 45$^\circ$.
To avoid charging of the surface a dual beam charge compensation scheme was applied consisting of electrons and low energy Ar$^+$ ions.

\subsection{Electrical Characterization}
For electrical and photo-dependent measurements, contacted \ce{MoS2} samples placed in a custom-built vacuum chamber. Measurements were conducted through electric feedthroughs with a two-channel Keithley 2612B source meter. All measurements were done under vacuum conditions ($\leq5.5\times10^{-5}$~mbar). For photodependent measurements, the channel was illuminated with a 532~nm laser at 0.1~mW laser power focused through a 10x objective from Zeiss (NA 0.25).

\subsection{Ion Irradiation}
Ion irradiation followed a standard protocol \cite{Zheng.2026}. An ion sputtering gun IG35/IG70 from OCI Vacuum Microengineering Inc. was used. Argon gas (99.999\% purity; Air Liquide)  was filled into the chamber, with a base pressure of $p_\text{base}\approx1\times10^{-8}$~mbar, until a pressure of $5\times10^{-7}$~mbar was reached. A Wien filter for mass separation is employed. The ion current was calibrated to be $0.36$~$\mu$A/cm$^2$. The devices were exposed to the beam until the desired fluence was reached.

\subsection{DFT Calculations}
%PK
Spin-polarized electronic structure 
%First-principles 
calculations were performed using Density Functional Theory (DFT) with VASP 6.4.2 version package \cite{Kresse.1999,Kresse.1996}. The interactions between ions and valence electrons were described using the projector-augmented wave (PAW)\cite{Blochl.1994} method, and electronic exchange and correlation were treated by the generalized gradient approximation of Perdew, Burke, and Ernzerhof (PBE)\cite{Perdew.1996}. The monolayer \ce{MoS2} models were constructed in a $5\times  5$ supercell. We employed a plane wave cut-off energy of 500~eV and an additional correction to the van der Waals dispersion interaction in the form of D3 according to Grimme \textit{et al.} \cite{Grimme.2011}, employing a Becke-Johnson damping function. 25 \AA\: of vacuum was applied to avoid interaction with the periodic images. A $2\times  2\times 1$ Monkhorst-Pack k-point grid was used to sample the 2D Brillouin zone. The structures were relaxed with PBE until the energy was less than 0.01 eV/~\AA{} while the self-consistent convergence limit was $10^{-6}$~eV. The unfolded band structure was plotted using the assistance of the VASPKIT code \cite{Wang.2021b}. The chemical potentials ($\mu_i$) used to determine the defect formation energies were referenced to their respective elemental reservoirs. Specifically, bulk Mo, bulk S, and graphite (C) were utilized as solid reservoirs, with their reference energies defined as the total energy of the bulk supercell divided by the number of constituent atoms in that supercell. For hydrogen, an isolated gaseous $\text{H}_2$ molecule was employed as the chemical potential reservoir. 

\subsection*{Author Contributions}
L.D., S.S., and M.S. and jointly conceived the idea of the study. L.D., under the supervision of M.S., designed the experimental research strategy, wrote the main content of the manuscript, conducted ion irradiation, conducted the electrical characterization, conducted Raman and PL measurements, and performed the associated data analysis. D.S., supervised by P.K., performed all theoretical calculations and co-wrote the manuscript. L.K., supervised by M.S., assisted with measurements and data analysis. O.K., supervised by M.S., synthesized and processed all samples. U.H. performed XPS measurements and the associated data analysis. O.A. performed AFM measurements. L.B. designed the electrical characterization setup. All authors
contributed to the interpretation of the results and to the preparation of the manuscript. M.S. supervised the overall project.

\subsection*{Acknowledgments}
The authors gratefully acknowledge financial support from the DFG within the IRTG 2803: 2D Mature, project No. 461605777 and project No. 429784087. Additionally, the authors acknowledge the computing time granted by the Center for Computational Sciences and Simulation (CCSS) of the Universität of Duisburg-Essen and provided on the supercomputer amplitUDE (DFG project 459398823; grant ID INST 20876/423-1 FUGG) at the Zentrum für Informations- und Mediendienste (ZIM). The authors acknowledge support by technical staff, especially Anke Hierzenberger, and helpful discussions with Bruno Schuler.

\subsection*{Financial disclosure}

None reported.

\subsection*{Conflict of interest}
The authors declare no potential conflict of interests.

\subsection*{Data Availability}

The data that support the findings of this study will be openly available following an embargo at the following DOI: 10.17172/nomad.rs84-nxz5
%\subsection*{Supporting information}
%AFM measurements after lift-off and cleaning; time-resolved gate-dependent measurements; changes in mobility and doping with progressing fluence; XPS-spectra of processed \ce{MoS2}; transfer characteristics of a \ce{MoS2} device and reference device for different irradiation steps with a logarithmic y-axis

\bibliographystyle{unsrt}   
\bibliography{wileyNJD-AMA}

\end{document}